# Mg$^{2+}$ ION EFFECT ON STABILITY OF DOUBLE-STRANDED POLYNUCLEOTIDE FORMED BY POLYRIBOINOSINIC AND POLYRIBOCYTIDILIC CHAINS

V.A. Sorokin, V.A. Valeev, E.L. Usenko, V.A. Karachevtsev

*Department of Molecular Biophysics, B. Verkin Institute for Low Temperature Physics and Engineering of NAS of Ukraine, 47 Nauky ave., 61103, Kharkiv, Ukraine*
*e-mail: usenko@ilt.kharkov.ua*

Mg$^{2+}$ effect on the thermal stability of the double-stranded polynucleotide polyI•polyC (IC) under conditions close to physiological ones (0.1M Na$^+$, pH7) was studied by the differential UV spectroscopy. The initial increase of the ion concentration ([Mg$^{2+}$]) leads to the increase of the melting temperature ($T_m$) and to the decrease of the absorption hyperchromicity value that is caused by the helix-coil transition ($h_{max}$). Mg$^{2+}$ interacts to π-electrons of hypoxanthine or cytosine rings (cation-π-interaction). When [Mg$^{2+}$] reaches a critical value of about 10$^{-4}$M, $T_m$ decreases and $h_{max}$ increase to values close to one observed in the absence of magnesium. It is supposed that at [Mg$^{2+}$]=[Mg$^{2+}$]$_{cr}$ IC transits into a new structural state. With [Mg$^{2+}$]>[Mg$^{2+}$]$_{cr}$ $T_m$ increases again and $h_{max}$ slowly decreases.

## I. INTRODUCTION

Problems concerning the elucidation of effect of divalent metal ion (Mt$^{2+}$) on the biological polynecleotides are caused by the ecological problems [1,2], the use of oligo- and polynucleotides metal complexes upon the creation of nanoelectronic schemes [3]. It generates a need for studying dependences of the conformational state of polynucleotides of different contents on Mt$^{2+}$ concentrations and temperature.

By the present time the above issues have been investigated in considerable detail for metal complexes of native DNA and double-stranded homopolynucleotides containing pairs of canonical nitrogen bases (AU and GC) [4-6]. Of some interest is to get similar information on polynucleotides including minor bases, in particular, hypoxanthine (H). This base, its nucleoside (inosine (Ino)) and nucleotide (inosinemonophosphate (IMP)) exhibit biological functions too. So, Ino is an active compound of some cardiologic preparations (for example, ribocsin). In replication or transcription processes GC-pairs can be replaced with IC-ones, without essential changes in structural parameters of the helix. An random substitution can lead to point mutations. The application of the targeted one allows using oligo- and polynucleotides containing IMP for solving problems of genetic engineering [7]. It should be noted too that this nucleotide is included in the anticodon of some t-RNA (for example, valine) [8]. And finally, the double-stranded polynucleotide polyI·polyC (IC) is used for the treatment of certain viral diseases as a potent agent that induces the interferon formation [8].

Earlier [6] we obtained a diagram of the conformational equilibrium for $Mg^{2+}$ ion complexes with polyA·polyU (AU). The present work was aimed at getting similar information for $Mg^{2+}$ ion complexes with IC the structure of which is significantly isomorphic to that of the canonic double helix of AU [8].

## II. MATERIALS AND METHODS

Potassium salts of polyriboinosinic (polyI) and polyribocytidilic (polyC) acids (Sigma Chemical Company). Double crystallized salt $MgCl_2·6H_2O$ (Reachim). Double-stranded IC was prepared by dissolution of polyI and polyC (with the 1:1 ratio) in 0.001M cacodylate buffer, pH7, containing 0.099M NaCl. As the self-association rate is proportional to the square of polynucleotide concentrations, to provide the most complete reaction of the IC formation, the solution the summing concentration of polynucleotides phosphorus (P) of which makes up 0.008÷0.01M was kept for 120 hours at 4°C. Before measurements the IC solution was dissolved up to the concentration $P=(3÷5)·10^{-5}M$ at room temperature ($T_0=22±1\ ^0C$). Such a small polymer concentration minimized aggregation between macromolecules in the presence of high concentrations of $Mg^{2+}$ ions. So, no noticeable light scattering on aggregates was observed up to 0.007M $Mg^{2+}$ both at room and maximum ($92^0C$) temperatures. The phosphorus concentration of IC was determined by the molar extinction coefficient $\varepsilon=4800\ M^{-1}\ cm^{-1}$ at $\nu=38500\ cm^{-1}$ [9]. The concentration of $Mg^{2+}$ ion added into the solution before measurements was determined by weight and controlled additionally by the complexonometric method. An error of the evaluation of the $Mg^{2+}$ ion concentration value did not exceed 0.5%, the concentration of the polynucleotide phosphorus being -1.5% [9,10].

The dependence of changes in the optical density of polynucleotide solutions (ΔA) on temperature (melting curves) as well as the dependence of ΔA on the wave number (ν) at various contents of $Mg^{2+}$ ions were measured at a Specord UV VIS spectrophotometer (Carl Zeiss Jena, Germany). The last was controlled with a personal computer which records polynucleotide melting curves as temperature dependences of the hyperchromic coefficient $h(T)=[\Delta A(T)/A_{T_o}]_{\nu m}$ ($A_{T_o}$ is the optical density of solution at $T=T_o$) and calculates derivatives $dh(T)/dT$. Melting curves of IC were registered at the wavenumber $\nu_m=40300\ cm^{-1}$ corresponding to the maximum change of polynucleotide absorption upon heating (Fig. 1). Their registration was performed by the differential scheme: identical solutions of polynucleotides or their complexes with $Mg^{2+}$ ions were in both the spectrophotometer channels. The reference cuvette was thermostated within ±0.5 $^0$C, the working one was heated to 92$^0$C, the heating rate being 0.25 deg/min. At this temperature the spectrum $\Delta A_h(\nu)$ was measured, corresponding to a change in IC absorption spectrum, induced with the transition of the ordered helix into the completely disordered coiled state. These changes correspond maximum values of the hyperchromic coefficient $h(\nu)_{max}$. $Mg^{2+}$ ion-induced change in IC absorption spectra were measured at $T=T_0$ by the four-cuvette scheme.

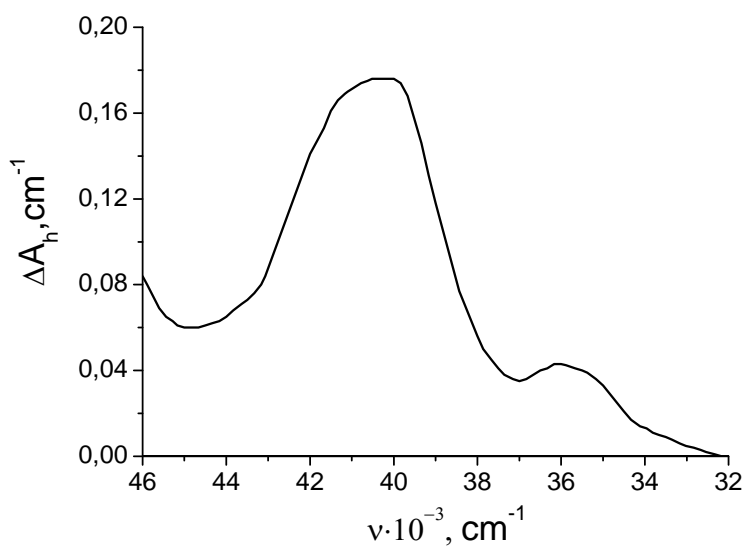

Fig. 1. Change in absorption spectrum of polyI·polyC, induced with heating from 20$^0$C to 92$^0$C. The form of the spectra $\Delta A_h(\nu)$ does not depend on the $Mg^{2+}$ ion concentration.

### III. EXPERIMENTAL RESULTS AND DISCUSSION

Within the whole range of wavenumbers measured (46000÷32000 cm$^{-1}$) at $T=T_0$ $Mg^{2+}$ ions do not change practically the IC absorption spectrum (changes being within 0.003 o.u.), that points to no presence of the ion formation of inner-sphere coordination bonds with heteroatoms of IC hypoxanthine and cytosine rings. Upon ionization of the hydroxyl group of ribose of CMP its UV spectrum must change too [11]. Thus, $Mg^{2+}$ ions do not interact with the oxygen atom of the ribose ring of the IC polycytidylic strand too that is natural for solution with pH7 as pK of CMP ribose ionization is about 13 [11].

The form of polynucleotides melting curves is a sensitive indicator of their conformation at room temperature. So, as is seen from Fig.2a, the single-stranded polyI dissolved in 0.001M sodium cacodylate, after adding 0.099M Na$^+$ into solution and lowering its concentration to P=5·10$^{-5}$M, holds a conformation of the single-stranded coil for 90 min in any case. However, being kept in the solution containing 0.1M Na$^+$ for 1 month, poly I (P~0.01M) transits into the state of the four-stranded helix (Fig. 2a). At 0.1M Na$^+$ (P~10$^{-4}$M) its melting temperature $T_m$ ($T_m$=T at which the derivative dh(T)/dT is of the maximum value) is equal to 27$^0$C that agrees satisfactorily with $T_m$= 29 $^0$C obtained by Hovard and Mils [4]. Melting of the four-stranded helix is characterized with high cooperativity (dh(T)/dT = 0.16 deg$^{-1}$) and a relatively high value of the hyperchromic coefficient ($h_{max}$=0.5). IC melting proceeds (Fig. 2a) in a temperature range stretched by tens of degrees (dh(T)/dT = 0.002 deg$^{-1}$) and is followed with a significantly lower increase of absorption ($h_{max}$=0.13 – Fig. 2a). Low values of dh/dT and h result from the fact that at neutral pH and 0.1M Na$^+$ poly C is in a conformation of a partially melted single-stranded helix [12]. The formation of the double-stranded (or four-stranded) helix of poly C is possible only in acid solutions [4,12,13].

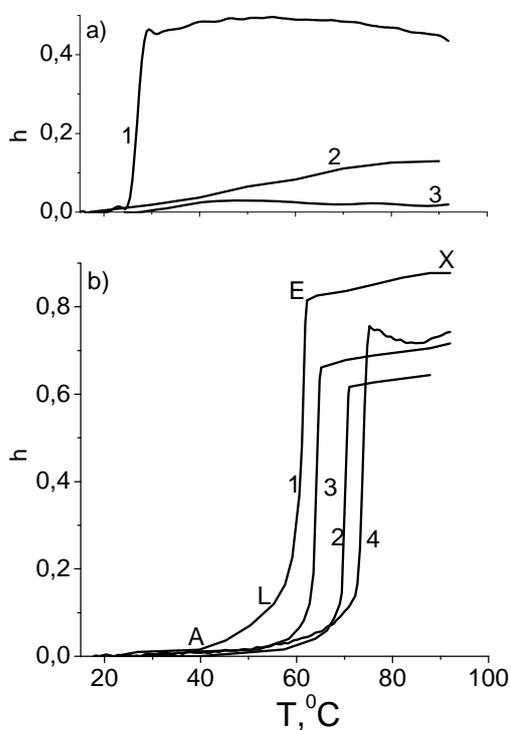

Fig. 2. Melting curves of polynucleotides in solution containing 0.1M Na$^+$ and Mg$^{2+}$ ions.
a) 1 – four-stranded poly I (keeping of single-stranded poly I for 1 month); 2 – single-stranded poly C; 3 - single-stranded poly I (keeping for 90 minutes). Melting curves of poly I and poly C were registered at $\nu$=40300 cm$^{-1}$ and $\nu$=37300 cm$^{-1}$, respectively. ([Mg$^{2+}$]=0). b) – Poly I· poly C: 1 - [Mg$^{2+}$]=0; 2 – 1.1·10$^{-4}$M Mg$^{2+}$. Poly I· poly I· poly C is suggested: 3 – 1.25·10$^{-4}$M; 4 – 0.007M Mg$^{2+}$.

Thus, upon the long-term keeping of the equimolar mixture of poly I and poly C both double IC helices and four-stranded 4·I helices can be formed [4,14]. Nevertheless, the formation of the last is likely not to take place in the range (25÷40), no absorption hyperchromicity is present, and the melting temperature ($T_{mo}$) of the polynucleotide formed is equal to 61 $^0$C (Fig. 2b). But upon preparing new mixtures of poly I and poly C rather great dispersion (±1.5 $^0$C) between $T_{mo}$ values was observed. Within this dispersion the mean $T_{mo}$ magnitude (61.5 $^0$C) coincides with $T_{mo}$=60 $^0$C obtained for poly I·poly C in [15]. The formation of IC upon mixing poly I and poly C is evidenced too by the agreement of the value $T_{mo}$=38.3 $^0$C obtained by us at 0.01M with $T_{mo}$=38 $^0$C determined for IC in [10]. $h_{max}$ values at $\nu$=38500 cm$^{-1}$ agree satisfactorily too: 0.29 (the present work) and 0.26 ([10]). A great $h_{max}$ value

(Fig. 2b) at $\nu=\nu_m$ points to a high helicity degree of the polynucleotide formed upon poly I and poly C mixing.

AU pairs in poly A· poly U and those in poly I· poly C are formed by two hydrogen bonds, and as it has been already noted these double helices are isostructural [8]. This probably conditions a low difference between IC and AU thermal stabilities (Fig. 2) (for AU at 0.1M Na$^+$ and 0.01M Na$^+$ $T_{mo}=(56\div58)$ $^0$C [15,16] and $T_{mo}=(37.5\div39)$ $^0$C [6,16], respectively). As well, the form of IC and AU melting curves is similar. In particular, as well as in AU, melting curves of IC break down into three main parts: the A-L section is probably a result of IC untwisting at the ends [6] or of melting of helical parts of single-stranded polyC having not formed a complementary pair with polyI; the L-E section characterized with the derivative $dh/dT \sim (0.2\div0.3)$ deg$^{-1}$ corresponds to the IC→I+C (2→1) transition; and the E-X section may be naturally attributed to melting of helical parts of polyC formed because of strand separation.

In the context of the above, it could be expected that $T_m$ dependences on Mg$^{2+}$ ion concentrations for AU [6] and IC are close. But for IC this dependence is of an unusual form (Fig. 3a).

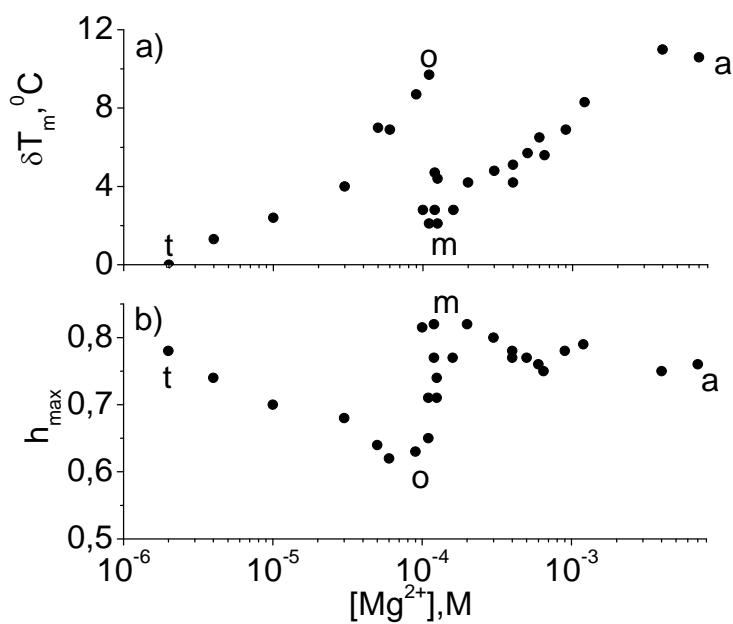

Fig. 3. Concentration dependences of change in melting temperature (a) and maximum hyperchromic coefficient at $\nu_m=40300$ cm$^{-1}$ (b).

As in the case of AU [6], the thermal stability of IC increases in the region of the low magnesium content (section t – o) but, when the Mg$^{2+}$ ion concentration reaches some "critical" value ([Mg$^{2+}$]$_{cr}$ = (1.1±0.1)·10$^{-4}$M), the IC melting temperature decreases ubruptly. Therefore, within the uncertainty limits of the $[Mg^{2+}]_{cr}$ value (region o – m in Fig. 3a) two melting temperatures differing in ~8 $^{0}$C correspond to one value of the magnesium concentration. At neutral pH triple helixes poly I· poly I· poly C may be formed [17]. It can be suggested that, as in the case of AU for which in the presence of ~5·10$^{-5}$M $Mg^{2+}$ the 2→3 transition begins upon heating [6], a similar transition (2(IC)→C2I+C) takes place for IC. At $[Mg^{2+}]>[Mg^{2+}]_{cr}$ the thermal stability of the polynucleotide formed increases again (Fig. 3a) and reaches its maximum value at ~0.003M $Mg^{2+}$. Perhaps, the section m – a (Fig. 3a) corresponds to a temperature change in the poly I· poly I· poly C→I+I+C (3→1) transition.

As is seen from Fig. 3b, correlation between changes in the melting temperature and in the $h_{max}$ value is observed. So, in the section t – o the $h_{max}$ value lowers but increases again in the narrow region of the ion content, corresponding to $[Mg^{2+}]\sim[Mg^{2+}]_{cr}$, reaching a value close to the $h_{max}$ at $[Mg^{2+}]$=0 ($h_{max}$ = 0.82±0.03). With the further rise of the ion concentration some decrease of $h_{max}$ is observed which is however within the maximum experimental dispersion of values for this parameter (Fig. 3b).

It is known that the Coulombic interactions of alkali and alkali-earth metal ions with negative charges on oxygen atoms of phosphate groups of single- and double-chain polynucleotides result in the increase of their thermal stability [4-6,16]. If melted links are present in macromolecules, this interaction induces their transition into the helical conformation and, as a result, the increase of the $h_{max}$ value. As the magnitude of $h_{max}$ is proportional to the initial helicity degree of the polynucleotide, the data from Fig. 3b permit to conclude that $Mg^{2+}$ ions cause its lowering in the region t – o. The cause of this anomaly is as follows. As analysis of the complexes structure of hydrated magnesium ions with DNA and RNA fragments shows, along with the formation of complexes with unit pairs of electrons, $Mg(H_2O)_6^{2+}$ ions interact with π-electronic system of rings of polynucleotide nitrogen bases [18]. For example, in the case of magnesium complexes with Dikerson decamer and dodecamer duplexes consisting mainly of sequences of GC-pairs and of magnesium complexes with an anticodon shoulder of yeast phenylalanylic t-RNA the cation-π interaction of $Mg(H_2O)_6^{2+}$ with cytosine takes place. It is essential that this interaction results in structural changes followed by stacking disruption [18]. Therefore, it can be assumed that this is just an effect which conditions the decrease of $h_{max}$ observed in our case. $^1$H-NMR spectroscopy data obtained (to tell the truth) for $Mn^{2+}$ ion complexes with IC do not contradict this assumption as they evidence that the most probable site of the magnesium location is the major groove of the IC double helix [9]. In this case the distance d between the $Mn^{2+}$ ion and hypoxanthine iminoproton forming a hydrogen bond with cytosine N3 makes up 4.7Å [9]. That is a necessary condition for initiation of cation-π interactions (d<5.2Å), which can weaken or disrupt this bond is met [18].

It is known that the formation of the inner-sphere coordination bond with heteroatoms of nitrogen bases leads to a shift of absorption bands and a change in their intensity, and, as a result, DUVS spectra appear, being of a corresponding form [4] which differs from that one conditioned with the polynucleotide disordering (Fig. 1). However, $Mg^{2+}$ cation-π interaction with DNA and RNA is outer-sphere ones, and this, probably, conditions no changes in absorption spectra of IC in the presence of $Mg^{2+}$ ions.

In accordance with [18], cation-π interactions destabilize the duplex. Nevertheless, in reality only stacking disruption is observed. It is unclear to what extent this fact effects the thermal stability as the sign and magnitude of change in $T_m$ are conditioned with a difference in constants of ion binding to double- and single-chain conformations of polynucleotides [6].

## IV. CONCLUSIONS

1. In the range of low concentrations $Mg^{2+}$ ions lower the magnitude of increasing of IC absorption, induced by the helix-coil transition on heating. This effect is supposed to result from base stacking disruption induced by the ion interaction with π-electrons of aromatic rings.

2. In the wide range of concentrations $Mg^{2+}$ ions increase the IC melting temperature. But some "critical" ion concentration exists at which the thermal stability of the polynucleotide lowers and the helicity degree increases. Probably, this is caused with the IC transition into a new conformation being likely a triple helix.

## ACKNOWLEDGEMENT

The study presented in this publication has been made possible due to the UB2-2442-KH-2 Grant of USA Civilian Research and Development Foundation (CRDF) for independent states of the former Soviet Union.